# The effect of premature wall yield on creep testing of strongly-flocculated suspensions

Anthony D Stickland[*,1], Ashish Kumar[1], Tiara E Kusuma[1], Peter J Scales[1], Amy Tindley[2], Simon Biggs[2] and Richard Buscall[1,2,3]

[1]Particulate Fluids Processing Centre, Department of Chemical and Biomolecular Engineering, The University of Melbourne, Victoria 3010, Australia

[2]Institute of Particle Science and Engineering, School of Process, Environmental and Materials Engineering, The University of Leeds, Leeds LS2 9JT, United Kingdom

[3]MSACT Research and Consulting, 34, Maritime Court, Haven Road, Exeter, EX2 8GP, United Kingdom

**Abstract**

Measuring yielding in cohesive suspensions is often hampered by slip at measurement surfaces. This paper presents creep data for strongly-flocculated suspensions obtained using vane-in-cup tools with differing cup-to-vane diameter ratios. The three suspensions were titania and alumina aggregated at their isoelectric points and polymer-flocculated alumina. The aim was to find the diameter ratio where slip or premature yielding at the cup wall had no effect on the transient behaviour. The large diameter ratio results showed readily understandable material behaviour comprising linear viscoelasticity at low stresses, strain-softening close to yielding, time-dependent yield across a range of stresses and then viscous flow. Tests in small ratio geometries however showed more complex responses. Effects attributed to the cup wall included delayed softening, slip, multiple yielding and stick-slip events, and unsteady flow. The conclusion was that cups have to be relatively large to eliminate wall artefacts. A diameter ratio of three was sufficient in practice, although the minimum ratio must be material dependent.

[*]Tel.: +61 3 8344 3430; Fax: +61 3 8344 4153
*Email address*: stad@unimelb.edu.au.

# 1 Introduction

The rheological properties of industrial suspensions such as mineral tailings, biosludges, pigments and ceramics, and household products including cosmetics, paints and foods, are often the major factor determining their use and applicability. In the industrial setting, a thorough understanding of suspension rheology allows for better prediction and optimisation of transport and handling processes, such as pipelining, mixing, solid-liquid separation and pumping. At some point, the suspension rheology limits the maximum particulate concentration that can be processed. In other cases, the flow properties of suspensions are formulated to produce a particular flow response such as when a toothpaste tube is squeezed – toothpaste yields and flows onto the toothbrush but stops flowing on cessation of the applied stress.

Flocculated suspensions, where the particles are attracted to each other, develop a networked structure capable of withstanding an applied load at solids concentrations above the gel point $\phi_g$ (Michaels and Bolger 1962; Rehbinder 1954). At sufficiently small loads or deformations, such suspensions behave as viscoelastic solids. If a shear load is large enough, the suspensions yield and flow viscously. $\phi_g$ can be much lower than random close packing – as low as 1 volume percent for biosludges (Stickland et al. 2008), for example – such that the yielding behaviour of strongly flocculated suspensions is important over a wide range of concentrations.

In shear, it is common to extract parameters from the flow curves for strongly-flocculated suspensions by fitting them to viscoplastic models such as Bingham (Bingham 1916) or Herschel-Bulkley (Herschel and Bulkley 1926), where the strength of the network is represented by the shear yield stress $\tau_y$, which is a function of the density and distribution of bonds in the system. The assumption embodied in these simple models and in yield stress as an operational parameter is that, at some critical stress, the material converts instantaneously from a perfectly rigid solid to a viscous liquid (Bird et al. 1982; Hartnett and Hu 1989; Liddell and Boger 1995). This is unlikely to be true in reality, even if it suffices to model many flow curves approximately. Hence, it comes as no surprise that experiments that probe yielding itself show the transition to be more progressive and to have an intrinsic time-dependence. For example, creep testing reveals viscoelasticity at stresses below yielding followed by time-dependent yield over a range of stresses rather than at a single stress (Barrie et al. 2004; Gibaud et al. 2010; Le Grand and Petekidis 2008; Uhlherr et al. 2005). Other descriptions of yield such as critical strain energy $W_y$ or yield strain $\gamma_y$ may be more appropriate for suspension behaviour (ten Brinke et al. 2008) compared to yield stress. For example, yield may be described by sharp (but not instantaneous) strain thinning or strain softening (Klein et al. 2007; Kobelev and Schweizer 2005; Kumar et al. 2012). In fact, as demonstrated here, an elastic body can yield without any need for an explicit yield criterion if the strain-softening of the modulus is strong enough.

The study of suspension viscoelastic and yielding properties and consequently the development of strongly-flocculated suspension constitutive descriptions cannot proceed without accurate measurement. Unfortunately, the accurate measurement of suspension properties is inherently difficult due to complex inter-particle and hydrodynamic interactions. The results from conventional rotational geometries such as cone-and-plate or cup-and-bob (Couette) are influenced by wall effects. Adhesive yielding allows depletion and slip (Buscall et al. 2009; Buscall et al. 1993). In addition, narrow gaps such as cone-and-plate are not suitable due to particle bridging (Boger 1999).

Since the 1980's, various authors (Barnes and Carnali 1990; Keentok et al. 1985; Nguyen and Boger 1985a) have substituted the concentric cylinder system of a Couette geometry with the vane-in-cup geometry (see in Figure 1) to alleviate the problem of slip at the surface of the inner cylinder surface in suspension rheology (Barnes 1995; Nguyen and Boger 1983). The vane also has the advantage of minimal structural disruption during testing (Barnes 1995; Barnes and Carnali 1990). The flow start up at controlled shear-rate vane technique (Nguyen and Boger 1983; 1985a) has been used in characterising a wide array of concentrated networked suspensions, including for example titania (Uhlherr et al. 2005; Zhou et al. 2001), alumina (Fisher et al. 2007; Johnson et al. 2000; Zhou et al. 2001; Zhou et al. 1999), brown coal (Leong et al. 1995) and wastewater sludges (Baudez et al. 2011).

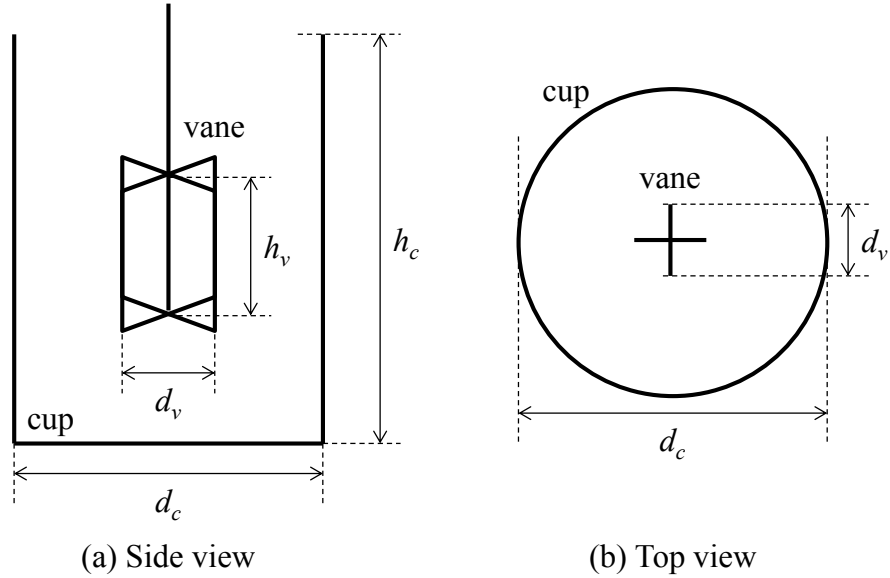

(a) Side view     (b) Top view

**Figure 1: (a) Side and (b) Top views of the four-bladed vane-in-cup geometry with a wide gap. $h_v$ and $d_v$ are the height and diameter of the vane while $h_c$ and $d_c$ are the height and diameter of the cup**

The vane replaces the inner cylinder in a concentric cylinder arrangement and thus alleviates slip at the inner surface. In a radial geometry, the stress and deformation decay with radial position in the gap relative to the vane diameter. In most cases, vanes are used in cup-to-vane diameter ratios of approximately 1.1 so that the deformation is approximately linear and the transformation of angle to strain is straightforward. However, since the wall yield stress of a suspension is typically less than the bulk yield stress (Buscall et al. 1993; Grenard et al. 2014; Saak et al. 2001; Walls et al. 2003), small diameter ratios are subject to slip at the cup wall. Whilst roughened surfaces have been employed with mixed success, the *vane-in-infinite-cup* geometry alleviates all wall effects and has been shown (Fisher et al. 2007) to be a suitable method for determining the yield stress and steady-state flow behaviour of strongly-flocculated particulate suspensions. However, the geometry has had limited use studying the transient elastic deformation of suspensions prior to yielding or the time dependencies of yielding and it is unclear what constitutes an infinite gap. Coussot and co-workers (Coussot et al. 2006) used a 25 mm diameter six-bladed vane in a 36 mm cup ($d_c/d_v$ = 1.44) to study the creep behaviour of a networked bentonite suspension (along with Maille mustard and Vivelle Dop hair gel as examples of a particulate suspension in an oil-in-water emulsion and a Carbopol gel respectively). They too justified using a larger than normal diameter ratio in order to minimise wall and edge effects; however, they did not vary the gap to show if indeed this had been achieved.

The work in this paper compared the creep behaviour of strongly-flocculated alumina and zirconia systems as functions of cup-to-vane diameter ratio, $d_c/d_v$. The expectation was that there would be a critical $d_c/d_v$ value above which the results would show no wall effects. A constitutive description when there is first wall yield and then true yield is complicated. The behaviours at the wall and in the bulk are both highly non-linear and hence it is difficult to posit a constitutive equation in advance, except in very general terms. We have used a power-law dependent model for the elastic modulus here, but in general, there is limited discussion of a constitutive description for suspensions. Whilst this is vitally important and the subject of current research (Buscall et al. 2014a; Buscall et al. 2014b; Lester et al. 2014), the aim of this work was to qualitatively compare the measured data to show when there was no influence of the cup wall. Development cannot proceed without data that is clear of wall artefacts, otherwise material behaviour is attributed that is in fact due to the measurement.

Two types of strongly-flocculated particulate mineral oxide suspensions were tested in this work. The term strongly-flocculated reflects that the strength of the inter-particle bonds was greater than about 20 $kT$ such that the suspensions were not influenced by Brownian motion, compared to weakly-flocculated suspensions that show long-term fluctuations and creep. The first type of suspension was either zirconia or alumina coagulated by van der Waals forces at their respective iso-electric points (IEPs). The second type was alumina flocculated with a high molecular weight polymer flocculant. The creep results from these suspensions were compared to highlight the susceptibility of different systems to apparent wall effects. In particular, the stresses where wall slip was observed in small diameter ratios were used to determine the ratio of bulk to wall yield stresses. In addition, the elastic deformation was analysed using an appropriate non-linear to show wall effects on suspension strain softening.

## 2    Theory

Most rheometer geometries use a relatively small gap so that the shear stress $\tau$ is approximately uniform and the shear strain $\gamma$ is easy to calculate (Barnes et al. 1989). Using the vane-in-infinite-cup geometry presents the problem of extracting stress-strain information from the test results. The vane-in-infinite-cup is the vane analogy of the wide-gap concentric cylinder geometry (Krieger and Maron 1952; 1954). The shear stress distribution as a function of radial position $r$ for concentric cylinders is given by setting the inertial and viscous terms of the momentum balance to zero:

$$\frac{\partial \tau}{\partial r} + \frac{2\tau}{r} = 0 \qquad \ldots(1)$$

Integrating and evaluating the shear stress at the vane, $\tau_v = \tau(r_v)$, gives:

$$\tau(r,t) = \tau_v \left(\frac{r_v}{r}\right)^2 \qquad \ldots(2)$$

Thus the shear stress diminishes with the radial position squared. The stress propagates to the walls of the container and the cup-to-vane diameter ratio determines the magnitude of the stress at the cup wall. If yielding at the wall is characterised by a critical wall yield stress $\tau_{y,wall}$ and yielding in the bulk by $\tau_{y,bulk}$, the critical diameter ratio at which premature yielding will be observed is given by the square-root of the ratio of the bulk and wall yield stresses (Buscall et al. 1993; Grenard et al. 2014; Saak et al. 2001; Walls et al. 2003):

$$\left(\frac{d_c}{d_v}\right)_{min} = \sqrt{\frac{\tau_{y,bulk}}{\tau_{y,wall}}} \qquad \ldots(3)$$

Thus, measurements in a relatively large gap to determine the bulk yield stress and in a relatively small gap to determine the wall yield stress would give an estimate of the minimum critical diameter ratio.

In the cylindrical geometry, the strain $\gamma$ at position $r$ and time $t$ is defined as

$$\gamma = -r \frac{\partial \theta}{\partial r} \qquad \ldots(4)$$

where $\theta(r,t)$ is the angular displacement. Since the stress varies as a function of radial position in the wide gap, a constitutive relationship is required to combine equations 2 and 4 in order to convert displacement (or rotational rate) into strain (or strain-rate).

If the stress and stress-rate are linear functions with respect to strain and strain-rate then the material is linear viscoelastic and the solution is relatively straight-forward. By way of example of the expected variation with gap width for a linear material, the Standard Linear Solid (Voigt form) (SLS) is solved under constant stress creep conditions. The SLS, also known as the Zener model, is a Hookean spring with instantaneous elasticity $G_i$ in series with a Kelvin-Voigt element with a retarded elasticity $G_r$ and retardation time, $t_r$. The constitutive equation for this arrangement is (Roylance 2001):

$$\frac{t_r}{G_i}\dot{\tau} + \left(\frac{1}{G_r} + \frac{1}{G_i}\right)\tau = t_r \dot{\gamma} + \gamma \qquad \ldots(5)$$

Under constant stress conditions ($\dot{\tau} = 0$, $\tau = \tau_0$) in Cartesian coordinates, the solution to Eq. 5 is:

$$\gamma(t) = \tau_0 \left[ \frac{1}{G_i} + \frac{1}{G_r} \left[ 1 - \exp\left(-\frac{t}{t_r}\right) \right] \right] \quad \ldots(6)$$

In radial coordinates with the vane rotating ($\tau(r_v,t) = \tau_v$, $\theta(r_v,t) = \theta_v$) and the cup static ($\theta(r_c,t) = 0$), the solution for $\theta_v(t)$ using Eqs 2, 4 and 5 is:

$$\theta_v(t) = \frac{\tau_v}{2} \left[ 1 - \left(\frac{r_v}{r_c}\right)^2 \right] \left[ \frac{1}{G_i} + \frac{1}{G_r} \left[ 1 - \exp\left(-\frac{t}{t_r}\right) \right] \right] \quad \ldots(7)$$

The relationship between the rotation of the vane in a finite medium and in an infinite medium $\theta_{v,\infty}$ is given by:

$$\frac{\theta_v}{\theta_{v,\infty}} = 1 - \left(\frac{r_v}{r_c}\right)^2 \quad \ldots(8)$$

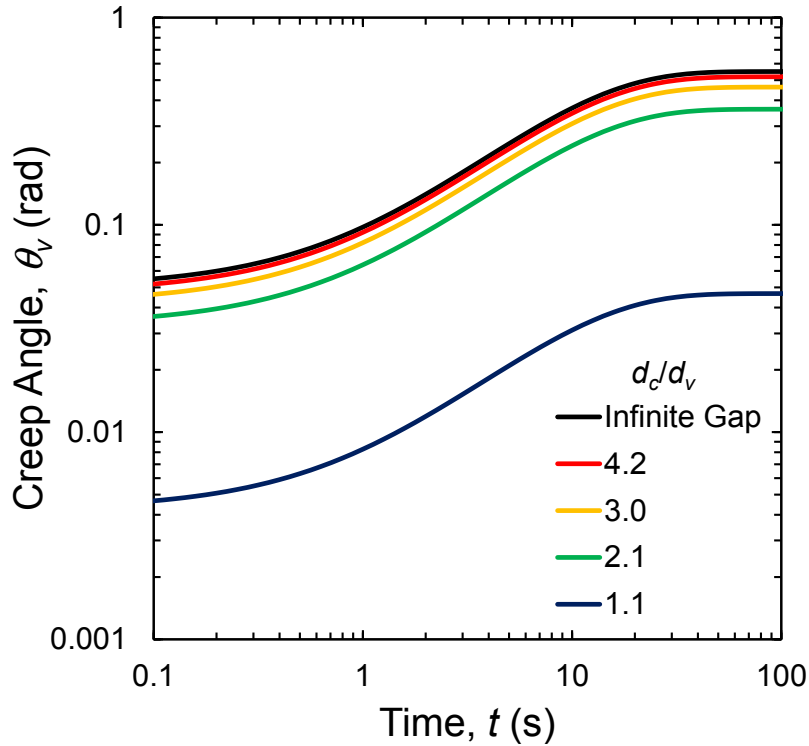

Figure 2: Variation of creep angle with time for various cup-to-vane diameter ratios for a Standard Linear Solid ($G_i$ = 1000 Pa, $G_r$ = 100 Pa, $t_r$ = 10 s, $\tau$ = 100 Pa)

Thus, for a linear material, diameter ratio does not affect the time dependency of the creep angle, just the magnitude. The creep angle $\theta_v(t)$ is plotted in Figure 2 using the cup-to-vane diameter ratios used in the alumina experiments for arbitrary material properties. The results show instantaneous compliance ($1/G_i$) followed by retarded compliance ($1/G_r$). For the smallest gap the creep angle is only 8% of the value for the infinite gap. As the gap increases, the creep angle asymptotes towards the infinite solution. The 4.2 cup-to-vane diameter ratio gives a creep angle of 94% of the infinite solution.

Comparing Eqs. (5) and (6) leads to the commonly used linear relationship between strain and angle of rotation at the vane (Krieger and Maron 1954):

$$\gamma_v = 2\theta_v \left[1 - \left(\frac{r_v}{r_c}\right)^2\right]^{-1} \qquad \ldots(9)$$

This relationship holds for all linear materials and in the small gap limit (when all materials are approximately linear). To complicate matters, suspensions are non-linear, such that the conversion between angle and strain is not straightforward. The steady-state angular displacement or rotational rate can be converted to steady-state strain and strain-rate respectively (Krieger and Maron 1952), but it is not possible to use the same method to convert time-dependent data for non-linear materials since their deformation history is stress and time dependent. For suspensions that yield, for example, it would be necessary to know the radial position within the gap at which the suspension is yielding in order to define the strain.

For creep data prior to yielding that is not time-dependent such as the initial or steady-state elastic deformation (that is, either none or all of the stresses have relaxed), it is possible to use a power-law approximation akin to Krieger and Maron to extract the non-linear strain and strain modulus. For a power-law dependent modulus $G(\gamma) = A\gamma^p$ (where $\tau = G(\gamma)\gamma$), the conversion from angle to strain is given by:

$$\gamma_v = \frac{2}{(1+n)}\theta_v \left[1 - \left(\frac{r_v}{r_c}\right)^{\frac{2}{1+n}}\right]^{-1} \qquad \ldots(10)$$

The linear approximation (Eq. 9) can therefore be used to determine the local slope of the power-law (calculated over 3 or 5 points, for example), which can then be used to calculate the real strain and modulus.

The results herein generally showed a low shear modulus $G_0$ followed by strain-softening that asymptoted to a constant power-law index. This motivated using a modified Cross equation (with modulus and strain instead of viscosity and strain-rate) to fit $G(\gamma)$:

$$G(\gamma) = \frac{G_0}{1 + \left(\frac{\gamma}{\gamma_y}\right)^{-n}} \qquad \ldots(11)$$

where $\gamma_y$ is the 'yield strain' when the modulus is ½$G_0$. This equation gives a linear material with modulus $G_0$ for $\gamma \ll \gamma_y$ and strain-softening varying with $\gamma^p$ for $\gamma \gg \gamma_y$. Note that at large strains and if $n \leq -1$ then the material softens as fast or faster than it can strain (the stress $\tau = G(\gamma)\gamma$ shows a maximum). If $n < -1$, the stress is non-monotonic with strain.

There is no explicit solution for stress as a function of strain based on Eq. 11 such that it is not possible to pose a direct analytical conversion from angle to strain. However, there is an explicit solution in the special case where $n = -1$. In this case, the modulus can be written as a function of stress:

$$G(\tau) = \frac{\tau}{\gamma} = G_0\left(1 - \frac{\tau}{\tau_y}\right) \qquad \ldots(12)$$

$G(\tau)$ asymptotes to a 'yield stress' $\tau_y$ where $\tau_y = G_0\gamma_y$. Integrating the combination of $G(\tau)$, the stress as a function of radius (Eq. 2) and the definition of strain (Eq. 4) from $\theta(r_v) = \theta_v(\tau_v)$ to $\theta(r_c) = 0$ gives:

$$\theta_v(\tau_v) = \frac{\tau_y}{2G_0}\left[\ln\left(1 - \frac{\tau_v}{\tau_y}\frac{r_v^2}{r_c^2}\right) - \ln\left(1 - \frac{\tau_v}{\tau_y}\right)\right] \qquad \ldots(13)$$

Thus, for materials that have a zero strain modulus $G_0$ and strain-soften with a slope of $n = -1$, Eq. 13 provides an analytical equation that can be fitted directly to experimental data to extract $G_0$ and $\tau_y$.

In order to get the conversion factor between $\gamma$ and $\theta$ for $n = -1$, simply substitute $\tau_v(\gamma_v) = G(\gamma_v)\gamma_v$ and rearrange in terms of $\gamma_v(\theta_v)$:

$$\gamma_v = \frac{\tau_y}{G_0}\left[\exp\left(\frac{2G_0\theta_v}{\tau_y}\right) - 1\right]\left(1 - \frac{r_v^2}{r_c^2}\right)^{-1} \qquad \ldots(14)$$

## 3    Materials and Methods

All reagents used were analytical grade and used without further purification. All reagents were prepared using Milli-Q water (Millipore™ Synergy®, 0.22 μm filter).

### 3.1    Coagulated Titania Suspensions

Rheological tests were performed using high purity anatase titania (ANX type-N, Degussa) in the laboratory at Leeds University. The purity and lack of surface contamination was verified using ion chromatography analysis (Thermo Dionex). The mean diameter $D$[4,3] estimated from low angle laser light scattering data (Malvern Mastersizer 2000) was 1.96 µm. The density was 3.78 g/cm$^3$. The IEP was measured (Colloidal Dynamics Zeta probe, Colloidal Dynamics Pty Ltd, Australia) as pH 6.5 (Biggs and Tindley 2007). The titania was prepared at a volume fraction of 0.153 by dispersion into 0.01 M KNO$_3$ at pH 3 using vigorous agitation by hand for 5 mins. The samples were pH adjusted and allowed to age for between 18 and 24 h to ensure chemical equilibrium. The pH was adjusted to the IEP and allowed to equilibrate prior to measurement.

### 3.2 Coagulated Alumina Suspensions

Rheological tests were performed using alumina (AKP-30, Sumitomo Chemicals Pty. Ltd, Japan) at The University of Melbourne. This system has been characterised in detail by Zhou *et al* (Zhou et al. 2001; Zhou et al. 1999). The alumina was nearly spherical with a density of 4.0 g/cm$^3$ and a mean diameter of 360 nm (Malvern Mastersizer 2000).

Coagulated alumina suspensions were prepared at a solids volume fraction of 0.20 in a 0.01 M KNO$_3$ background electrolyte solution. The mixture was acidified using 1 M HNO$_3$ to pH 5 (EZDO PL-600 pH meter with Sensorex Combination pH electrode, Denver Instrument Company), at which the suspension was fully dispersed. The mixture was then sonicated with a high intensity sonic probe (Misonix™ S4000 Sonicator, 12.5 mm horn, 20 kHz, 800 W) to ensure complete dispersion and then left to rest for 24 to 48 hours to establish physical and chemical equilibrium. Before measurements, the pH was adjusted using 1 M KOH to the isoelectric point (IEP) of approximately pH 9.2 (Foong 2008; Zhou 2000) and allowed to rest for a minimum of 2 hours. The sample had a nominal yield stress of 85 Pa, as measured using the Nguyen and Boger (Nguyen and Boger 1983; 1985a) vane technique.

### 3.3 Flocculated Alumina Suspensions

Polymer-flocculated systems are known to be irreversibly shear-history dependent. Therefore, an experimental protocol was implemented here that minimised shear degradation. Dispersed alumina suspensions were prepared at a volume fraction of 0.03 at pH 5 by dispersing the alumina powder in Milli-Q water and then sonicating, as above. Suspensions were then transferred to specially designed cylinders (see Figure 3) for flocculation.

The flocculant was an anionic, high MW polyacrylate / polyacrylamide copolymer flocculant (AN934SH, SNF Australia, 30% anionic). Stock polymer solutions were prepared at 2 g/L and agitated overnight on a rotary shaker. Stock bottles were covered with aluminium foil and kept refrigerated to minimise polymer degradation; unused stock was discarded after a week. An hour prior to use, the flocculant stock was diluted to 0.1 g/L and homogenised using a magnetic stirrer (Gladman et al. 2005).

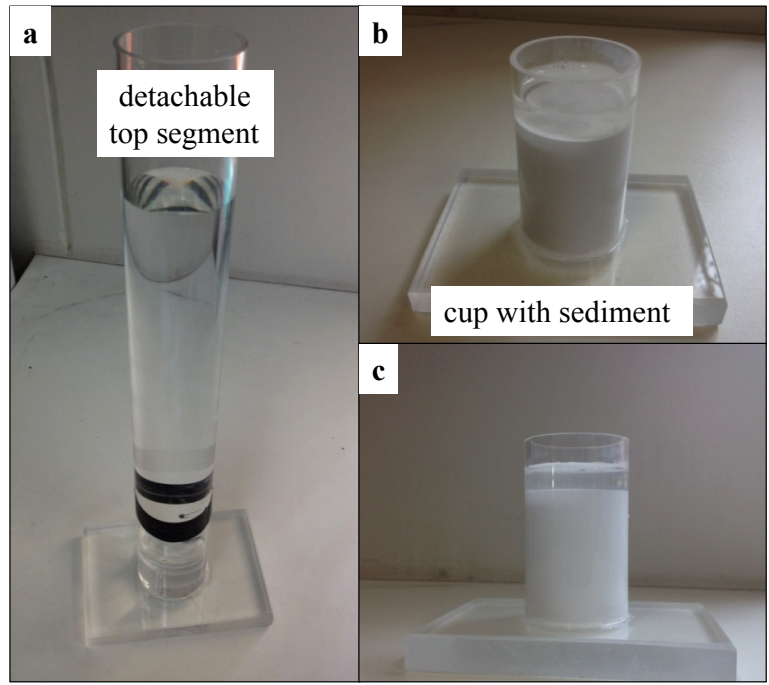

**Figure 3: (a) Perspex cylinder with detachable top segment. The detachment point is indicated by the black marks on the cylinder. (b and c) Upon flocculation and settling, the top segment was removed allowing subsequent rheological testing**

The polymer dose was 40 gram per tonne of suspended solids, which was chosen from preliminary flocculation tests. Higher dosages showed little or no improvement in turbidity or settling rate, indicating the dose of 40 g/t was an optimum dosage under the flocculation conditions. Polymer solutions were added directly to suspensions in the settling cylinders using the 'plunger' method. This was a simple, low-cost method of achieving consistent mixing and flocculation conditions (Gladman et al. 2005; Usher 2002). Upon addition of flocculant, a perforated disk attached to a long rod (the plunger) was manually moved up and down the cylinder at a constant rate (3 s per stroke) a fixed number of times (4 strokes).

The flocculated suspensions were allowed to settle for a minimum of 24 hours, or until no discernable change in sediment bed height was detectable. The cylinders had detachable top segments (see Figure 3(a)) such that, upon flocculation and settling, the supernatant could be removed with a syringe, the top segment removed and the remaining cup with sediment used in rheological testing (Jeldres et al. 2014). The average solids volume fraction of the settled sediment was determined by weight loss on drying to be 0.122 ± 0.010. This design limited exposure to shear. The only shear that the flocculated material would have experienced would have been during floc formation, during settling (van Deventer et al.

2011) and during insertion of the vane (although the vane does minimise sample disturbance compared to other rheometer geometries (Nguyen and Boger 1985b)). This experimental protocol was 'single-use' based on the premise that the flocculated samples were irreversibly changed by the rheological test.

*3.4  Rheology*

Constant stress creep experiments were performed using stress-controlled rheometers at Leeds University (C-VOR, Bohlin) and at The University of Melbourne (AR-G2, TA Instruments). The vane-in-cup geometry was used with a combination of vanes and cups to provide the range of $d_c/d_v$ ratios shown in Table 1. The tests on coagulated titania used two cup sizes with the same vane (combinations A and B), leading to cup-to-vane diameter ratios of 1.1 and 2.0 respectively. The creep tests were performed over a range of stresses for 50 s. The tests on coagulated alumina involved four different cup sizes with the same vane, leading to the ratios of 1.1, 2.1, 3.0 and 4.2 (combinations C to F). The polymer-flocculated tests involved the same cup (see Figure 3) with progressively smaller vanes (combinations G, H and I). The 10 mm vane was the smallest available, so a larger cup was used to give the ratio of 4.2 (combination J). The vanes were inserted into the middle of the sample with the same distance from the vane to the top of the sample and from the vane to the bottom of the cup.

During the creep test, a constant torque $T$ was applied and the creep angle, $\theta$, measured as a function of time. The torque was converted to shear stress using the relationship (Nguyen and Boger 1983):

$$T = \frac{\pi d_v^3}{2}\left(\frac{h_v}{d_v} + \frac{1}{3}\right)\tau \qquad \ldots(15)$$

Table 1. Vane-and-cup combinations with associated gap widths and cup-to-vane diameter ratios, $d_c/d_v$.

| Combination | Vane dimensions | | Cup dimensions | | Gap width (mm) | $d_c/d_v$ |
|---|---|---|---|---|---|---|
| | $h_v$ (mm) | $d_v$ (mm) | $h_v$ (mm) | $d_v$ (mm) | | |
| A | 30 | 14 | 47 | 15 | 1 | 1.1 |
| B | 30 | 14 | 60 | 28 | 7 | 2.0 |
| C | 42 | 28 | 80 | 30 | 1 | 1.1 |
| D | 42 | 28 | 105 | 60 | 16 | 2.1 |
| E | 42 | 28 | 95 | 85 | 28.5 | 3.0 |
| F | 42 | 28 | 200 | 118 | 45 | 4.2 |
| G | 42 | 28 | 78 | 32 | 2 | 1.1 |
| H | 50 | 15 | 78 | 32 | 8.5 | 2.1 |
| I | 40 | 10 | 78 | 32 | 11 | 3.2 |
| J | 40 | 10 | 80 | 41.5 | 15.8 | 4.2 |

In the case of coagulated titania, the samples were pre-sheared for 60 s at 100 s$^{-1}$ followed by a 300 s rest period. Between successive applied stresses, the coagulated alumina samples were sheared vigorously by hand with a spatula and then tapped on the bench to remove air bubbles before inserting the vane and waiting 20 minutes to reach thermal equilibration. Flocculated suspensions were replaced with new samples between each applied stress.

## 4    Results and Discussion

The creep results for the coagulated titania with two diameter ratios are presented first, followed by the results for coagulated and flocculated alumina in four different diameter ratios.

*4.1    Coagulated Titania*

Large Diameter Ratio ($d_c/d_v$ = 2.0)

The raw data of creep angle as a function of time for applied stresses between 30 and 80 Pa for the coagulated titania sample is presented in Figure 4(a), with the corresponding plot of rotational rate versus time shown in Figure 4(b) (the rotational rate was calculated from a 3-point running average over the angular displacement data). The coagulated titania in the 2.0 diameter ratio showed instantaneous elasticity convoluted with inertia (as signified by $\theta \propto t^2$) up to about 0.2 s, followed by retarded elastic deformation. The retarded elasticity became less pronounced at stresses closer to yielding (for example the 50 and 55 Pa angles only increased by 30% beyond 0.3 s compared with 160% at 30 Pa). At stresses less than 55 Pa, the samples did not yield within the experimental timeframe of 50 s, although there was some creeping flow at the higher two stresses (50 and 55 Pa) – this could have been wall slip perhaps, albeit much diminished in effect because of the wide gap. Above 60 Pa, the material showed instantaneous viscous flow. In the narrow stress range between 56 and 60 Pa, the material exhibited time-dependent yielding, with a maximum measured yield time of about 20 s at 56 Pa.

Small Diameter Ratio ($d_c/d_v$ = 1.1)

The coagulated titania was also tested using a small diameter ratio geometry as recommended by rheometer manufacturers. The vane diameter was the same as above ($d_v$ = 14 mm) but the cup was smaller ($d_c$ = 15.4 mm), leading to a cup-to-vane diameter ratio of 1.1. The creep angle and rotational rate results for the coagulated titania in the narrow gap are shown in Figure 5. A linear plot of angle is shown in Figure 6 for an alternative perspective.

Since the cup was not roughened, it was expected to see evidence of slip at the cup wall and it was immediately apparent that the small diameter ratio results were more complicated than the large ratio results.

- The initial inertia and instantaneous deformation were much reduced.
- At the lowest stress (30 Pa), the titania had solid-like retarded viscoelasticity, similar to the Standard Linear Solid behaviour in Figure 2.
- Next, at 40 to 46 Pa, the samples flowed at very slow rates, which was likely due to wall slip.

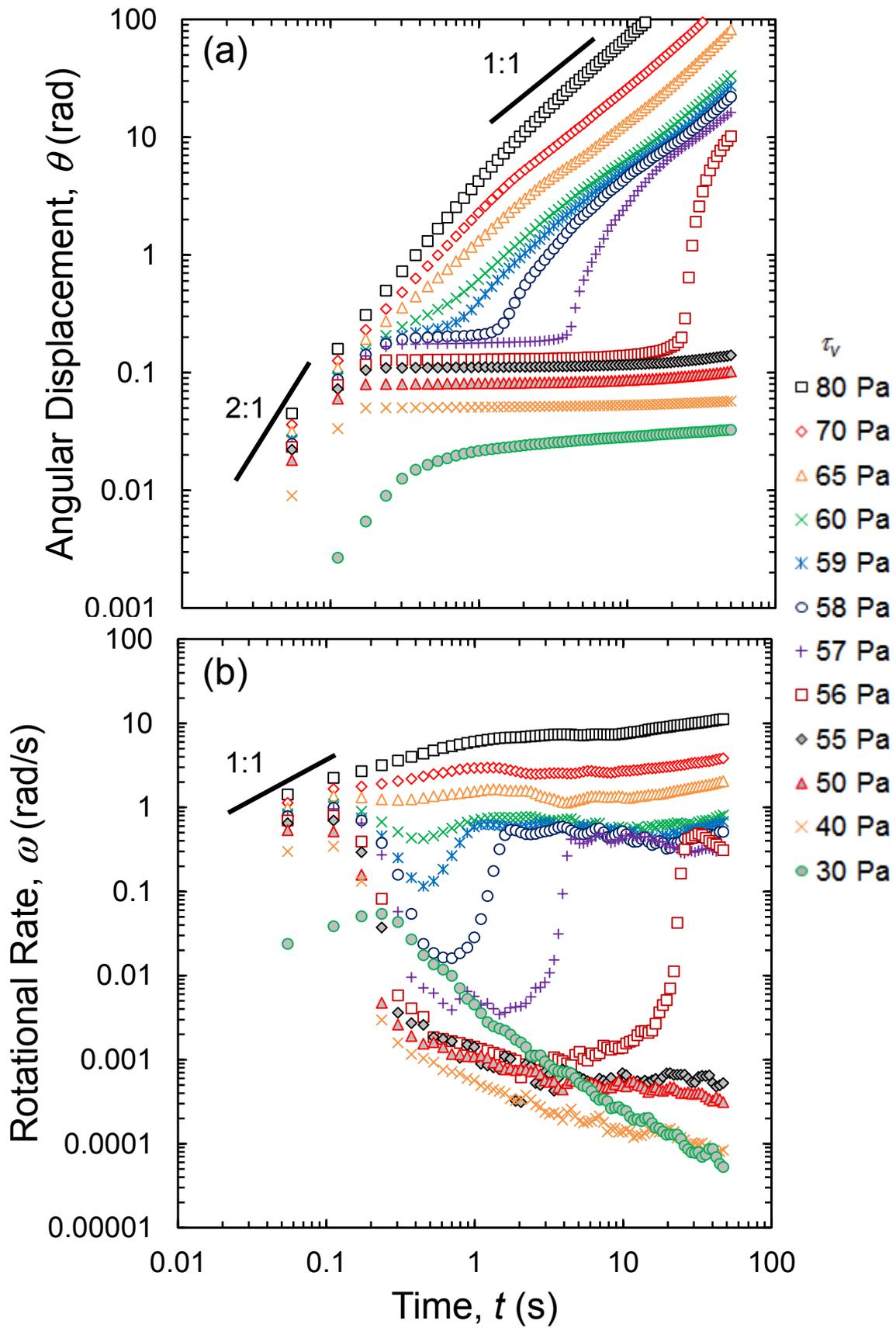

Figure 4: (a) Angular displacement and (b) Rotational rate as functions of time for creep testing of coagulated titania in a cup-to-vane diameter ratio of 2.0

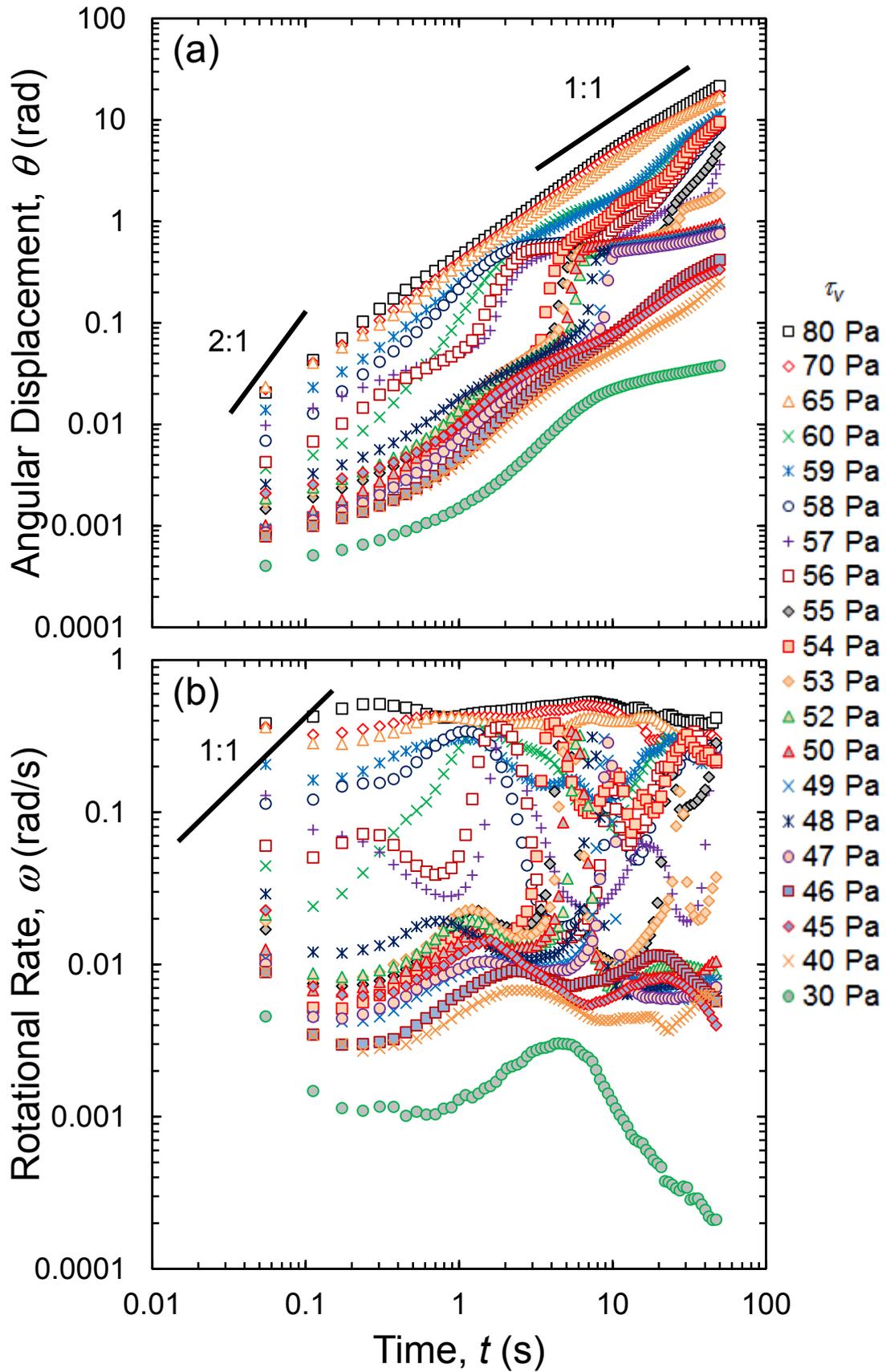

**Figure 5:** (a) Angular displacement and (b) Rotational rate as functions of time for creep testing of coagulated titania in a cup-to-vane diameter ratio of 1.1

- Above that (47 to 52 Pa), there were various stick-slip events combined with viscoelasticity. The slope at longer times was less than 1 on a logarithmic plot, indicating that steady-state viscous flow was not achieved.
- For stresses between 53 and 60 Pa, the coagulated titania initially showed retarded viscoelasticity before a sudden increase in displacement from about 0.08 rad to 0.5 rad at some time between 1 and 10 s. However, the material did not continue to flow, instead returning to retarded viscoelastic behaviour. The tests then showed subsequent yielding and stick-slip events. Presumably the suspension was yielding at the wall rather than in the bulk.
- For stresses 65 Pa and above, the suspension yielding almost instantaneously, although the rotational rate was not steady.

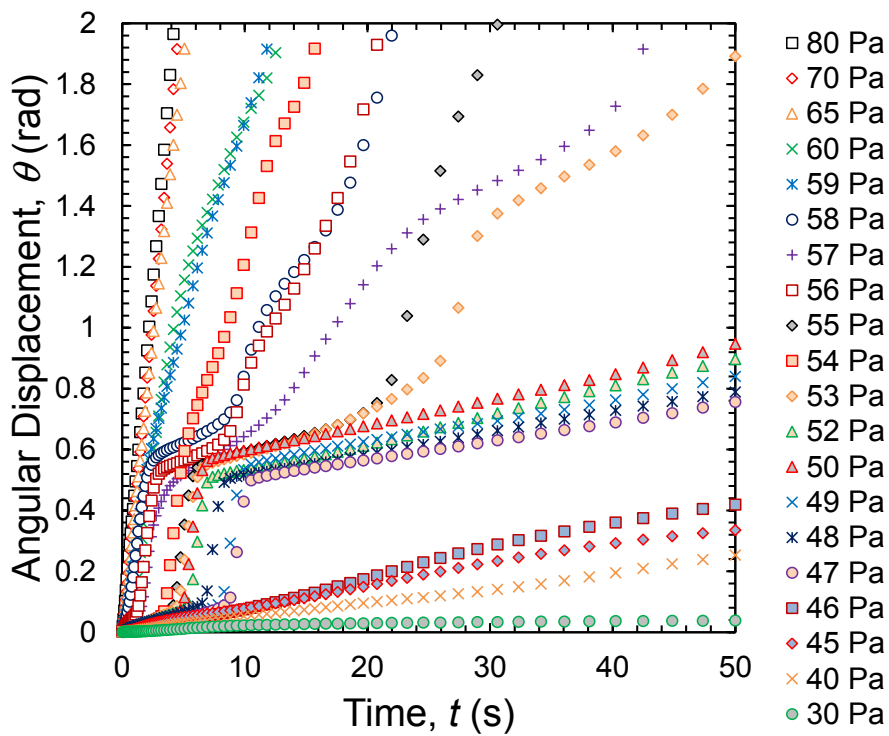

Figure 6: Linear plot of angular displacement as a function of time for creep testing of coagulated titania in a cup-to-vane diameter ratio of 1.1

Comparison between cup-to-vane diameter ratios

In order to account for the different geometries, the linear approximation given by Eq. 9 was used to calculate the apparent strain, which was divided by the stress to give the apparent compliance. A selection of results between 40 and 60 Pa are shown in Figure 7. There was stress dependence to the compliance such that the suspension rheology was non-linear at these stresses. Whilst this contradicted the use of the linear approximation, the linear conversion at least gave a first-order comparison of the narrow and wide gaps.

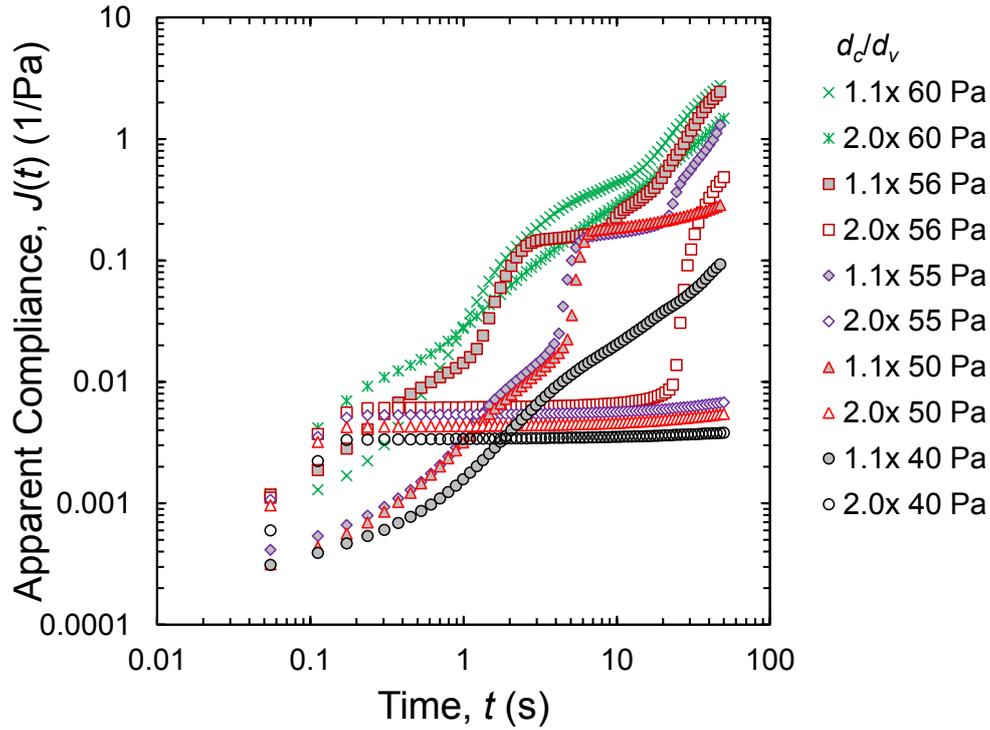

**Figure 7: Apparent compliance for a selection of creep stresses for coagulated titania in both small and large diameter ratios**

The large diameter ratio creep behaviour was relatively straightforward for this sample, with instantaneous deformation followed by retarded elasticity at low stresses (40, 50 and 55 Pa) and instantaneous and retarded elasticity followed by time dependent yield over a narrow stress range (56 to 60 Pa). The small diameter ratio creep behaviour was much more complicated with limited instantaneous elasticity, extended retarded viscoelasticity and long-term flow at low stresses (40 Pa). Stresses 47 Pa and above showed multiple yielding events followed by retarded deformation.

56 Pa was the lowest stress that showed yielding in the large cup, whereas the behaviour was consistent with yielding or slipping at the cup wall at 47 Pa vane stress for the small diameter ratio. Based on the $1/r^2$ dependency for stress, the wall stress in the narrow gap was 41 Pa for a vane stress of 47 Pa. Thus, the apparent adhesive to cohesive yield stress ratio was 0.73, which was consistent with the notion of wall yielding or adhesive failure occurring at significantly lower stresses than within the bulk (or cohesive failure) (Buscall et al. 1993; Grenard et al. 2014; Saak et al. 2001; Walls et al. 2003). Applying Eq. 3 gave a critical cup-to-vane diameter ratio of 1.17.

*4.2   Coagulated Alumina*

Creep testing of the coagulated alumina was performed using four different cup sizes over stresses ranging from 1 to 150 Pa.  The tests at 1 Pa were at the limit of sensitivity and were removed from the graphs here for the sake of clarity.  The apparent compliance was calculated from $J(t) = \gamma(t)/\tau$, where the apparent linear strain was given by Eq. 10.  Due to a higher data collection rate compared to the previous coagulated titania results, the rotational rate was calculated using a running average over 11 points.   Rotational data for early time creep ringing at low stresses have been removed.

<u>$d_c/d_v$ = 4.2</u>

The apparent compliance and rotational rate results for coagulated alumina in the largest diameter ratio are presented in Figure 8(a) and (b) respectively.

- At the lower stresses from 5 to 70 Pa, the results initially showed creep ringing (up to about 3 s for).  The ringing was followed by retarded viscoelastic behaviour.  The results at 5, 10 and 25 Pa were very similar, suggesting linear behaviour (that is, no stress dependence of the compliance).  Non-linearity, through softening of the retarded elasticity, began to appear at 40 Pa.
- At higher stresses but still below yielding (85 to 113 Pa), the samples showed greater instantaneous deformation followed by gradual retarded deformation, with the creep angle roughly doubling compared to orders of ten times bigger at lower stresses.  The instantaneous deformation was convoluted with inertia, as signified by $J \propto t^2$ and $\omega \propto t$, which extended to about 0.05 s.  Interestingly, the 110 and 113 Pa tests looked like they were about to yield between 100 to 200 s, showing an increase in rate, but they did not yield and continued to show solid-like creep.  This may have been due to rate-dependent hardening of the suspension at decreasing strain-rates (Buscall et al. 2014a).
- At 115 Pa and greater, the results showed instantaneous deformation, some retarded deformation and then time-dependent yield to steady-state flow.  By 150 Pa, the yield was essentially instantaneous.  The rotational rate showed a gradual increase over the experimental timeframe and did not reach steady-state within 1000 s.

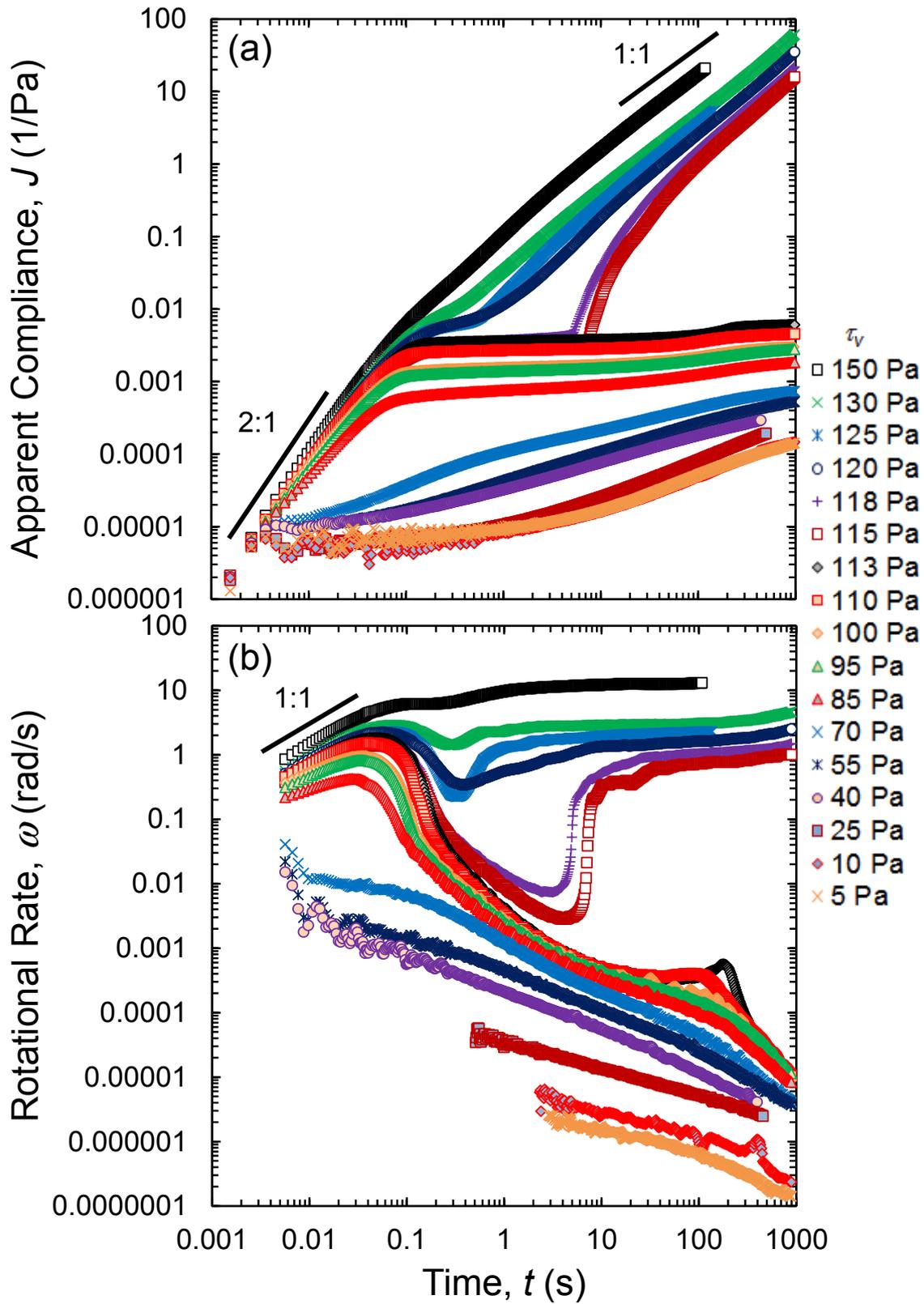

Figure 8: (a) Apparent compliance and (b) Rotational rate as functions of time for creep testing of coagulated alumina in a wide gap ($d_c/d_v$ = 4.2)

$d_c/d_v = 3.0$

The creep behaviour in the 3.0x diameter ratio was very similar to the 4.2x ratio and is therefore not presented for brevity. The same non-linear softening began to appear at 40 Pa. Time-dependent yield was observed at 113 Pa compared with 115 Pa in 4.2x ratio. There was no evidence of slip.

$d_c/d_v = 2.1$

The results for coagulated alumina in a cup-to-vane diameter ratio of 2.1 began to show notable differences compared to the 3.0x and 4.2x ratios (see Figure 9).

- The low stress results from 5 to 85 Pa showed the same creep ringing and retarded viscoelasticity as the wider gaps. However, the 2.1x ratio results were linear to higher stresses (up to 55 Pa).
- The 85 and 100 Pa results for the 2.1x ratio did not show the same increased instantaneous deformation and softening of the retardation as observed in the larger diameter ratios.
- At 115, 120 and 130 Pa, the suspension continued to show significant retarded deformation prior to time-dependent yielding. The retarded deformation was similar to inertia with $J \propto t^2$, but only after an initial delay.

$d_c/d_v = 1.1$

The apparent compliance and rotational rate results for the coagulated alumina in the smallest cup-to-vane diameter ratio are shown in Figure 10(a) and (b) respectively. The tests included two repeat measurements at 110 and 100 Pa.

- The low stresses (5 to 85 Pa) showed creep ringing (up to 3 s for 5 Pa) followed by retarded viscoelasticity. The results started to become non-linear at 40 Pa.
- The tests at 95 Pa, 97 Pa and 100 Pa (2) exhibited similar behaviour to 100 Pa for the 2.1x ratio with reduced instantaneous deformation compared to the wider gaps.

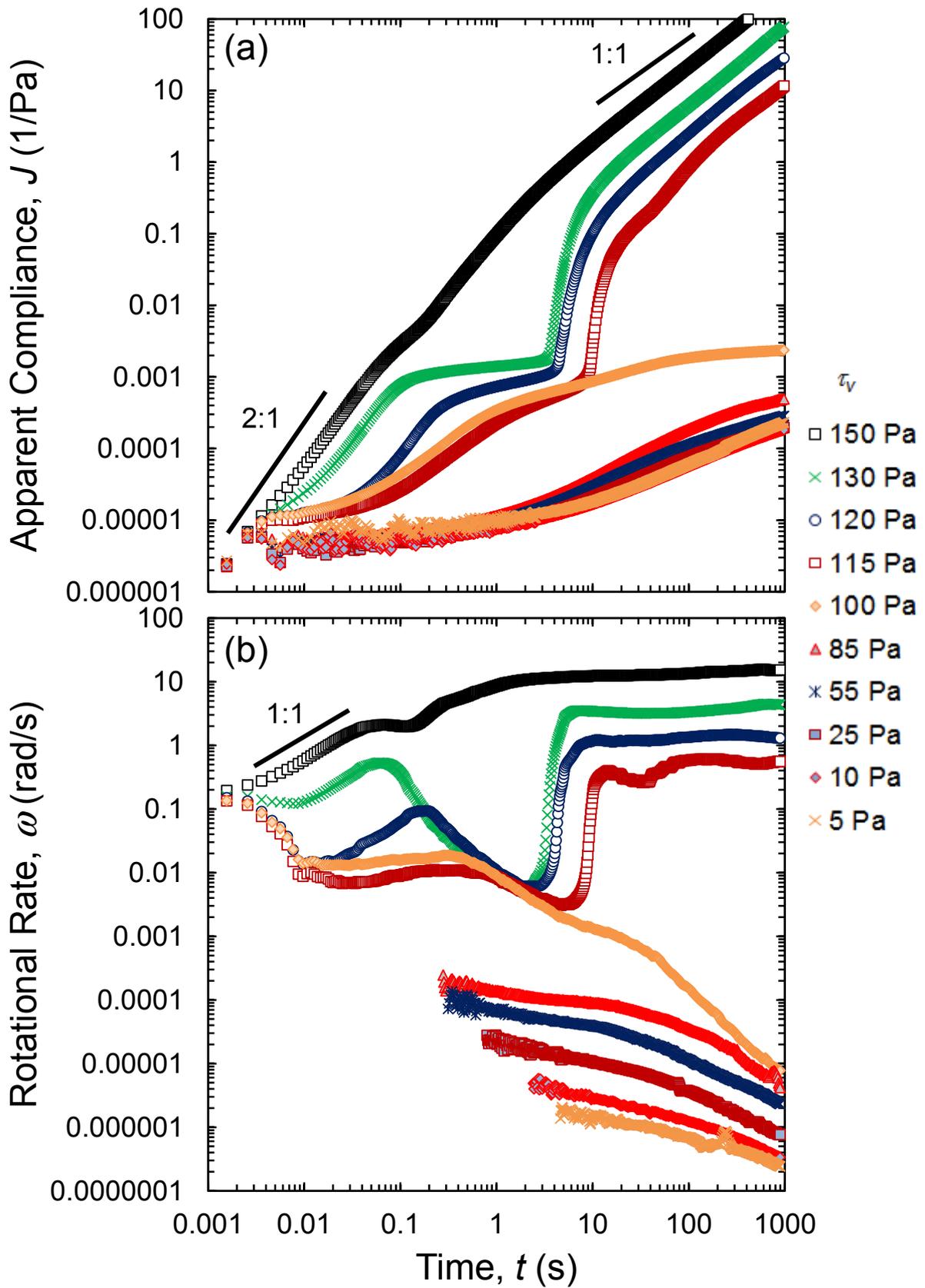

Figure 9: (a) Apparent compliance and (b) Rotational rate as functions of time for creep testing of coagulated alumina in a wide gap ($d_c/d_v$ = 2.1)

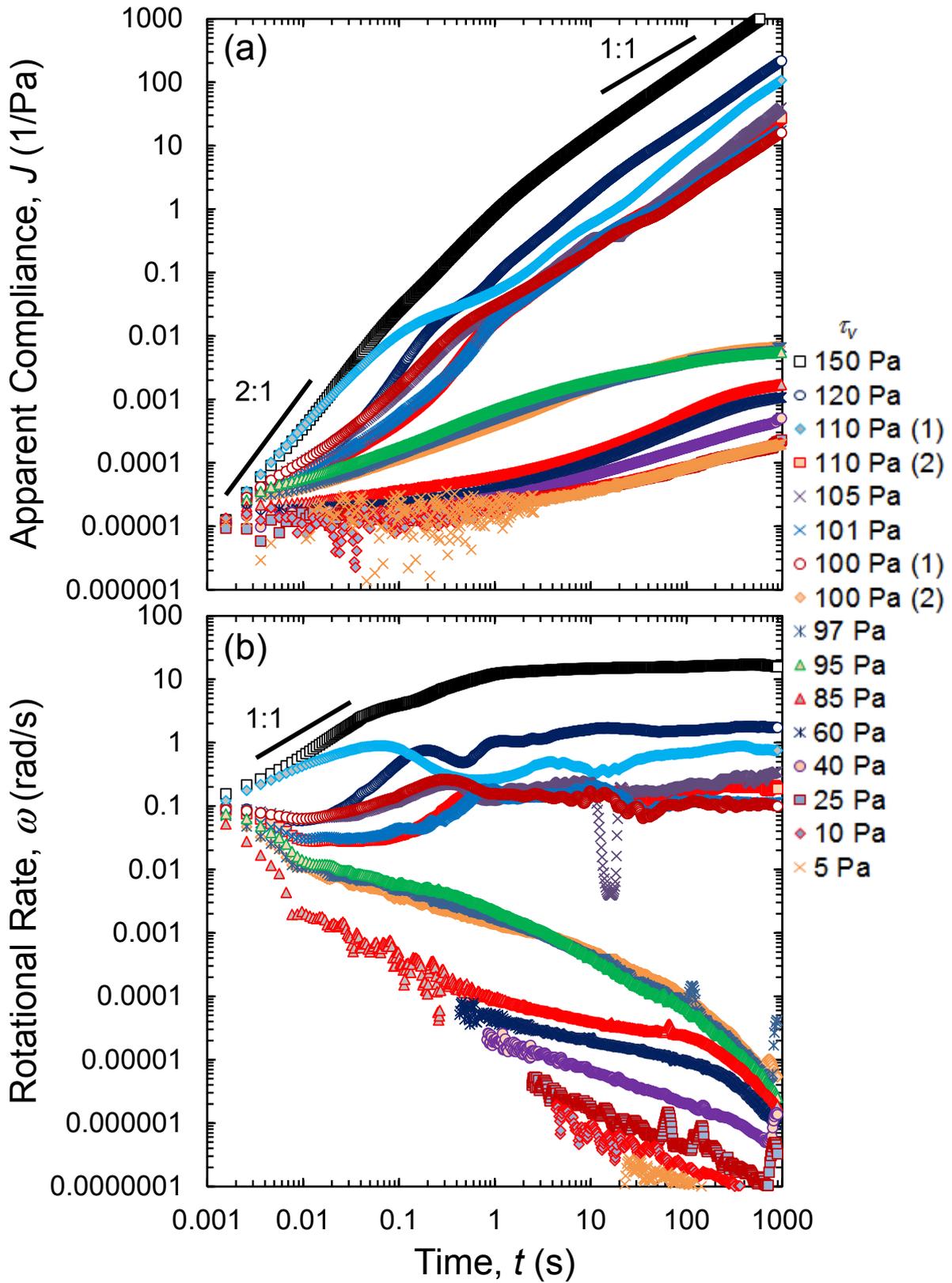

Figure 10: Apparent compliance as a function of time for creep testing of coagulated alumina in a narrow gap ($d_c/d_v$ = 1.1). (1) and (2) refer to repeats at the same stress

- The material yielded at 100 Pa and above. This was significantly lower than the 113 – 115 Pa measured in the larger diameter ratios and consistent with slip at the cup wall. The transient behaviour was also phenomenologically different to the wider gaps. There were no clear yielding events, rather reduced instantaneous deformation followed by un-steady flow. The delayed deformation suggested that slip had to occur at the cup wall before the suspension could flow. The rate for 105 Pa appeared to jam briefly at around 20 s, which could have been a wall artefact. Only the 120 Pa and 150 Pa tests reached steady-state.

It was difficult to get reproducible measurements in the small diameter ratio geometry, as evidenced by the repeated experiments at 100 and 110 Pa. The sample yielded for one test at 100 Pa (1) but not for the other (2). Likewise, the two tests at 110 Pa showed variable behaviour. This is in contrast to the larger ratio results that showed excellent progression through the stresses. All possible precautions were made to minimise experimental variability including concentration changes due to evaporation.

The coagulated alumina in the smaller two diameter ratios showed the same delayed deformation seen in the narrow gap results for the coagulated titania, although there was not the same dramatic oscillation in the rotational rate upon slipping. There were only small indications of the stick-slip behaviour in the alumina data that were so pronounced in the titania results. Likewise the low stress results did not show long term creep, although this may have been the short time scale of the titania experiments. In both cases, the cup wall was clearly influencing the suspension behaviour before and upon yielding.

Analysis of Coagulated Alumina Data

Given a wall yield stress of 100 x $(14/15)^2$ = 87 Pa and a bulk yield stress of 115 Pa, the ratio of the adhesive to cohesive yield stresses was 0.76 for the coagulated alumina. The critical cup-to-vane diameter ratio given by Eq. 3 was 1.15. This was similar to the 1.17 for the coagulated titania. However, one of the trends identified in the coagulated alumina data was that the 2.1x ratio had significantly reduced instantaneous deformation at stresses closer to yielding, suggesting the cup wall was not only affecting the yielding behaviour, but also the viscoelastic behaviour prior to yielding.

To illustrate, the creep angles at the start and finish of the experiments were extracted from the results. The early time, chosen to represent the instantaneous deformation $\theta_i$, was selected as 0.1 s, a time where the quadratic dependency due to inertia was diminished. The angle at 1000 s was assumed to be the steady-state creep angle $\theta_\infty$. The time-dependency of the creep tests was characterised by a retardation time, $t_r$, which was determined empirically from the 'half-life' $t_{1/2}$ taken to reach half the deformation between the instantaneous and steady-state deformations, i.e. $t_r = t_{1/2}/\ln(2)$, where $\theta(t_{1/2}) = \frac{1}{2}(\theta_i + \theta_\infty)$.

The apparent linear strain and moduli were calculated using the linear approximation (Eq. 9). A running 3 point average power-law index was then calculated, allowing the real strain and moduli to be calculated using Eq. 10. This was then fitted using the modified Cross model (Eq. 11), giving $G_0$, $\gamma_y$ (or $\tau_y$) and $n$. In order to then demonstrate the accuracy of the extracted parameters, the model was used in a finite difference numerical scheme to re-predict the creep angle.

For the smaller two diameter ratios, the analysis gave a modulus with an asymptotic index of close to -1 ($\pm$ 0.05). In these cases, it was much simpler to fit the exact solution for creep angle as a function of stress for $n = -1$ (see Eq. 13) and directly extract $G_0$ and $\gamma_y$ (or $\tau_y$).

The results for the instantaneous deformation are shown in Figure 11 with the modified Cross model parameters given in Table 2. The creep angle data was initially linear with stress, as illustrated by the dotted lines. The fitted zero strain moduli $G_{0,i}$ varied from 90 to 270 kPa, although the low strain data were somewhat suspect due to experimental sensitivity and creep ringing. $G_i$ was not expected to increase with strain for the 2.1x ratio, for example.

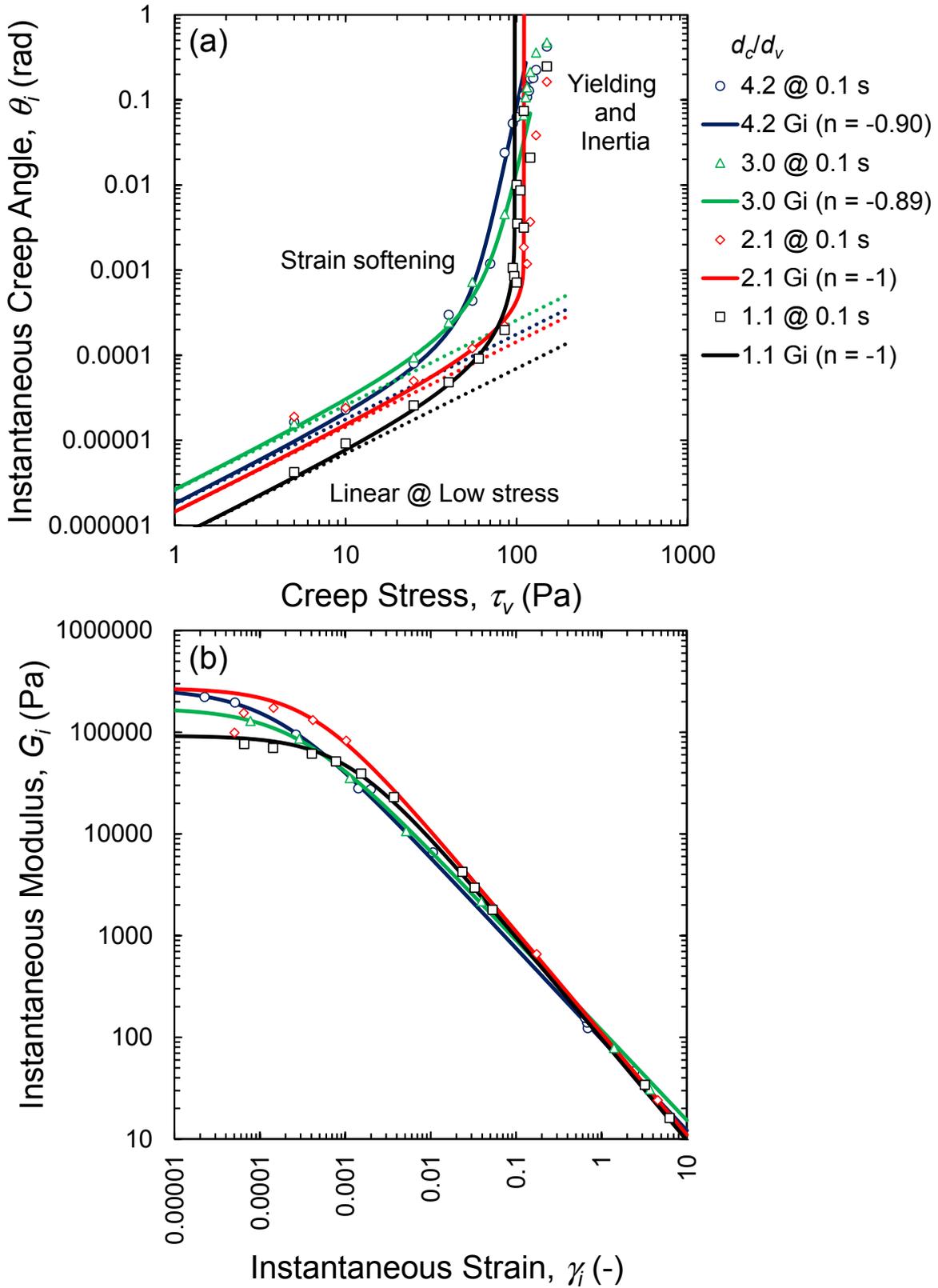

Figure 11: (a) Instantaneous creep angle $\theta_i = \theta(0.1\ s)$ as a function of stress and (b) instantaneous modulus as a function of instantaneous strain for coagulated alumina in various cup-to-vane diameter ratios. The modified Cross model had either variable $n$ or fixed $n = -1$.

Table 2: Modified Cross model fit parameters (see Eq. 11) for instantaneous modulus as a function of instantaneous strain for coagulated alumina in various cup-to-vane diameter ratios

| $d_c/d_v$ | $G_{0,i}$ (Pa) | $\gamma_{y,i}$ | $\tau_{y,i}$ (Pa) | $n_i$ |
|---|---|---|---|---|
| **1.1** | 92224 | $1.05 \times 10^{-3}$ | 97.1 | -1 |
| **2.1** | 271648 | $4.05 \times 10^{-4}$ | 110 | -1 |
| **3.0** | 172774 | $2.69 \times 10^{-4}$ | 46.5 | -0.887 |
| **4.2** | 267516 | $1.42 \times 10^{-4}$ | 38.1 | -0.896 |

As the strain increased, inter-particle bonds started being broken rather than just stretched or compressed and the modulus began to decrease. The onset of strain softening differed for the 4.2x and 3.0x diameter ratios compared to the 2.1x and 1.1x ratios. The 4.2x and 3.0x ratios showed critical strains about 0.0002 (corresponding to stresses about 40 Pa) followed by asymptotic power-law behaviour with an index of -0.9, whereas the 1.1x and 2.1x data began softening at higher stresses and strains with a strain softening slope of -1. In the softening region, the smaller two diameter ratios appeared stiffer with a higher modulus than the larger ratios up until a strain of 1.

This material yielded at strains of order 1, indicating 'cage melting' or steric hindrance as the yielding mechanism (Koumakis and Petekidis 2011; Pham et al. 2008) compared with bond breaking, which would lead to yielding at much lower strains. Strains greater than 1 were typical for the highest stresses that had presumably yielded instantaneously (< 0.1 s) and the deformation was no longer governed by just the material elasticity; the data in these cases were not included in the model fitting.

The suspensions deformed viscoelastically with time. The steady-state deformation results taken at 1000 s are shown in Figure 12 with the model parameters given in Table 3. The steady-state modulus showed similar trends to the instantaneous modulus, with the smaller diameter ratios having a higher modulus than the larger ratios up until strains of about 1. The strain softening indices for the larger diameter ratios were -0.83 and -0.9 whereas the smaller ratios were still at -1.

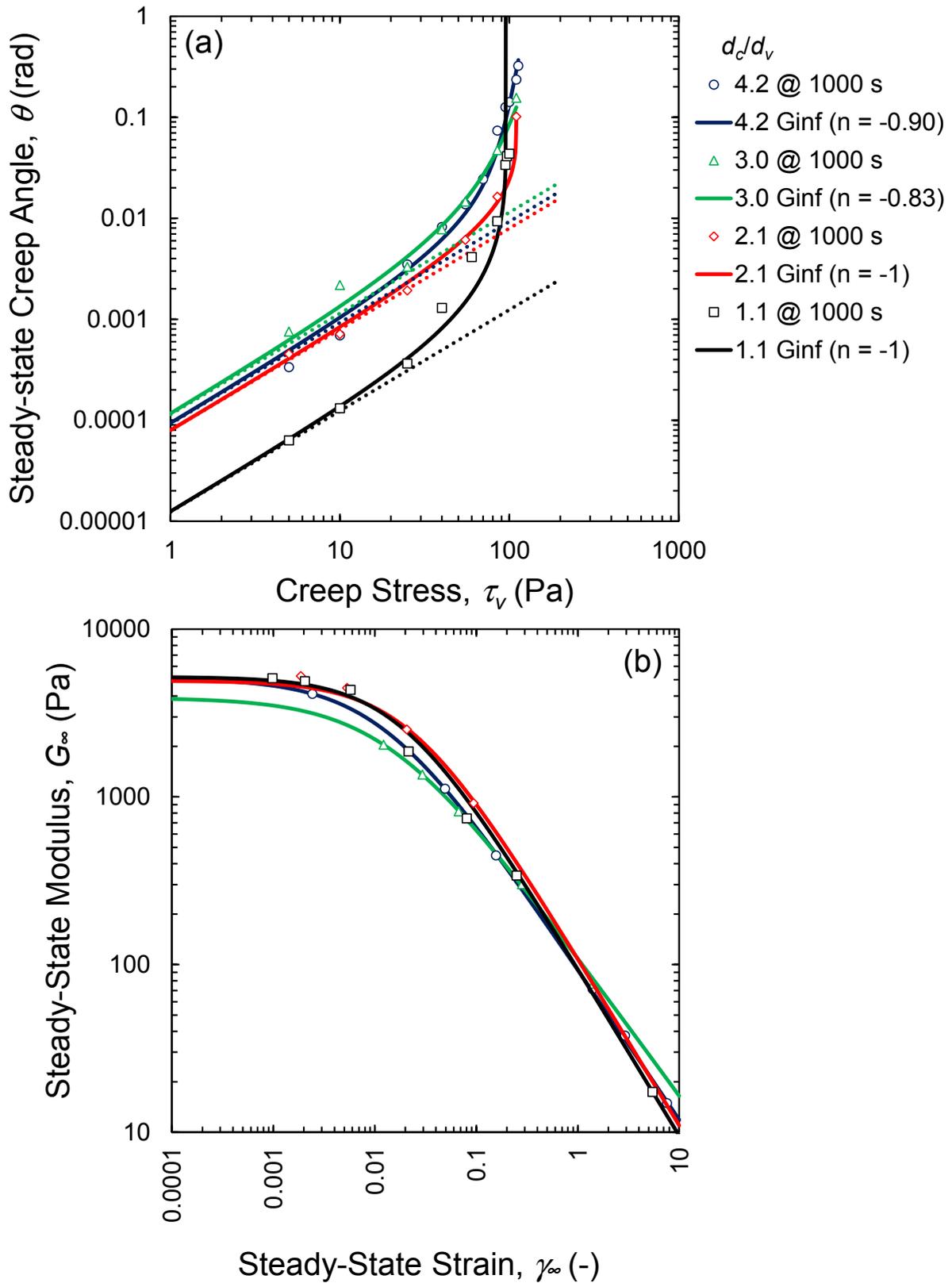

Figure 12: (a) Steady-state creep angle $\theta_\infty = \theta(1000\ s)$ as a function of stress and (b) steady-state modulus as a function of steady-state strain for coagulated alumina in various cup-to-vane diameter ratios. The modified Cross model had either variable $n$ or fixed $n = -1$

Table 3: Modified Cross model fit parameters (see Eq. 11) for steady-state modulus as a function of steady-state strain for coagulated alumina in various cup-to-vane diameter ratios

| $d_c/d_v$ | $G_{0,\infty}$ (Pa) | $\gamma_{y,\infty}$ | $\tau_{y,\infty}$ (Pa) | $n_\infty$ |
|---|---|---|---|---|
| **1.1** | 5199 | $1.83 \times 10^{-2}$ | 95.3 | -1 |
| **2.1** | 4928 | $2.23 \times 10^{-2}$ | 110 | -1 |
| **3.0** | 3902 | $1.37 \times 10^{-2}$ | 53.6 | -0.829 |
| **4.2** | 5091 | $1.20 \times 10^{-2}$ | 60.9 | -0.900 |

Given that the moduli were non-linear, the retardation time was also expected to vary with stress. The retardation times (see Figure 13) decreased from as much as 450 s at low stresses (although some of the very low stresses did not reach steady-state and the calculated retardation time was a lower estimate) to as little as 35 s at stresses close to yielding. The smaller (1.1x and 2.1x) and larger (3.0x and 4.2x) diameter ratios began to diverge at 55 Pa, which coincided with the point when the larger diameter ratios showed the onset of softening in the modulus results. Calculating the true retardation time would require a history-dependent constitutive model – the data here suggested a decaying power-law or exponential may be appropriate.

The onset of softening and the value of the softening index are important for understanding how suspensions yield. Using the data for the small diameter ratios of 1.1x and 2.1x at stresses close to yielding to give constitutive parameters would get the wrong material properties and the wrong interpretation of behaviour. Whilst the softening exponents were within the range observed for suspensions (which can vary from -0.7 (Pham et al. 2008) to -1.4 (Yin and Solomon 2008), for example), the outcome that the softening behaviour differed with diameter ratio confirmed that the phenomenological observation of different creep behaviours was not just due to non-linear affects but that the wall was preventing the deformation of the suspension. In other words, the suspension needed to soften at the wall in the small diameter ratios.

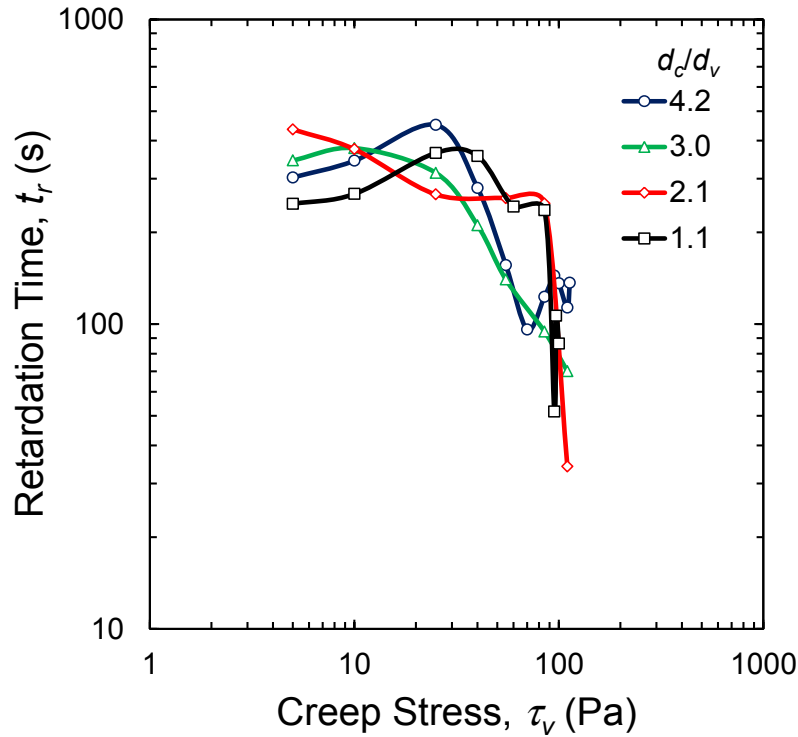

**Figure 13: Retardation times for coagulated alumina in various cup-to-vane diameter ratios (calculated from the half-life of the decay from 0.1 s to 1000 s)**

*4.3    Polymer-Flocculated Alumina*

Creep tests were performed on polymer-flocculated alumina in order to investigate if a different category of strongly-flocculated suspension behaved in a similar manner as the 'pH' coagulated systems.  A separate settling experiment was performed to prepare the sample in each case to minimise shear degradation.  The tests were performed at four different cup-to-vane diameter ratios at stresses of 10, 50 and 80 Pa, with each test requiring a separate batch settling experiment due to the shear-sensitive network.

The angular displacement results are shown in Figure 14.  Although the 80 Pa tests did pass relatively large angles, which suggested they were close to yielding, the stresses were all below yielding.  The tests showed inertia and creep ringing in some cases, then retarded viscoelastic behaviour.  A difference between the cup-to-vane diameter ratios emerged at 80 Pa.  The two larger diameter ratios showed significant instantaneous deformation followed by long-term creep, whereas the 1.1x and 2.1x diameter ratios did not show the same increase in instantaneous deformation.  The ratios of the creep angles at 1000 s and 0.1 s are shown in Figure 15, along with the retardation time as calculated from the half-life.  The larger diameter ratios had smaller relative retardation.  For example, the angle ratio at 80 Pain the 4.2x case was 9.97 whereas it was 105 in the 2.1x test, more than an order of magnitude different. The retardation time generally decreased with increasing stress, with shorter times for the larger diameter ratios at 80 Pa. There was an anomalous result that the 3.0x angular displacement was greater than for the 4.2x, which may be a reflection of the difficulty of preparing polymers solutions at the same state of ageing.

With only three values for each diameter ratio, analysis using the modified Cross model was difficult and very subjective. Therefore, the analysis was limited to the simpler linear approximation. The apparent linear moduli were calculated as a function of strain from the angles at 0.1 s and 1000 s (see Figure 16). The results for all the diameter ratios except 1.1x were in good agreement, except $G_i$ at 80 Pa for 2.1x. The 1.1x diameter ratio had higher instantaneous moduli, suggesting that it was stiffer than it should have been. The slopes at larger strains were $G_i \propto \gamma^{-0.877}$ and $G_\infty \propto \gamma^{-0.806}$, although there was not enough data to conclude whether these values were asymptotic or whether there was a difference between the diameter ratios. The steady-state strains at 80 Pa approached order 1, which suggested polymer-flocculated systems also had 'cage-melting' yielding.

Thus, the limited data for the polymer-flocculated alumina showed some of the same trends as the coagulated alumina sample at stresses below yielding, with increased instantaneous deformation followed by faster and relatively smaller retarded deformation for the larger diameter ratios. This indicates that wall effects on pre-yield viscoelasticity may manifest for all strongly-flocculated suspensions.

It was noted that each creep test on the flocculated alumina settled slightly by the end of the test (< 1 mm). Since the suspension had already settled to equilibrium, the additional settling after the application of shear stress indicated synaereses of the flocculated network, similar to the shear-densification seen in flocs at concentrations below the gel point (van Deventer et al. 2011). The shear stress was insufficient to yield the suspension in shear but large enough to locally densify the network.

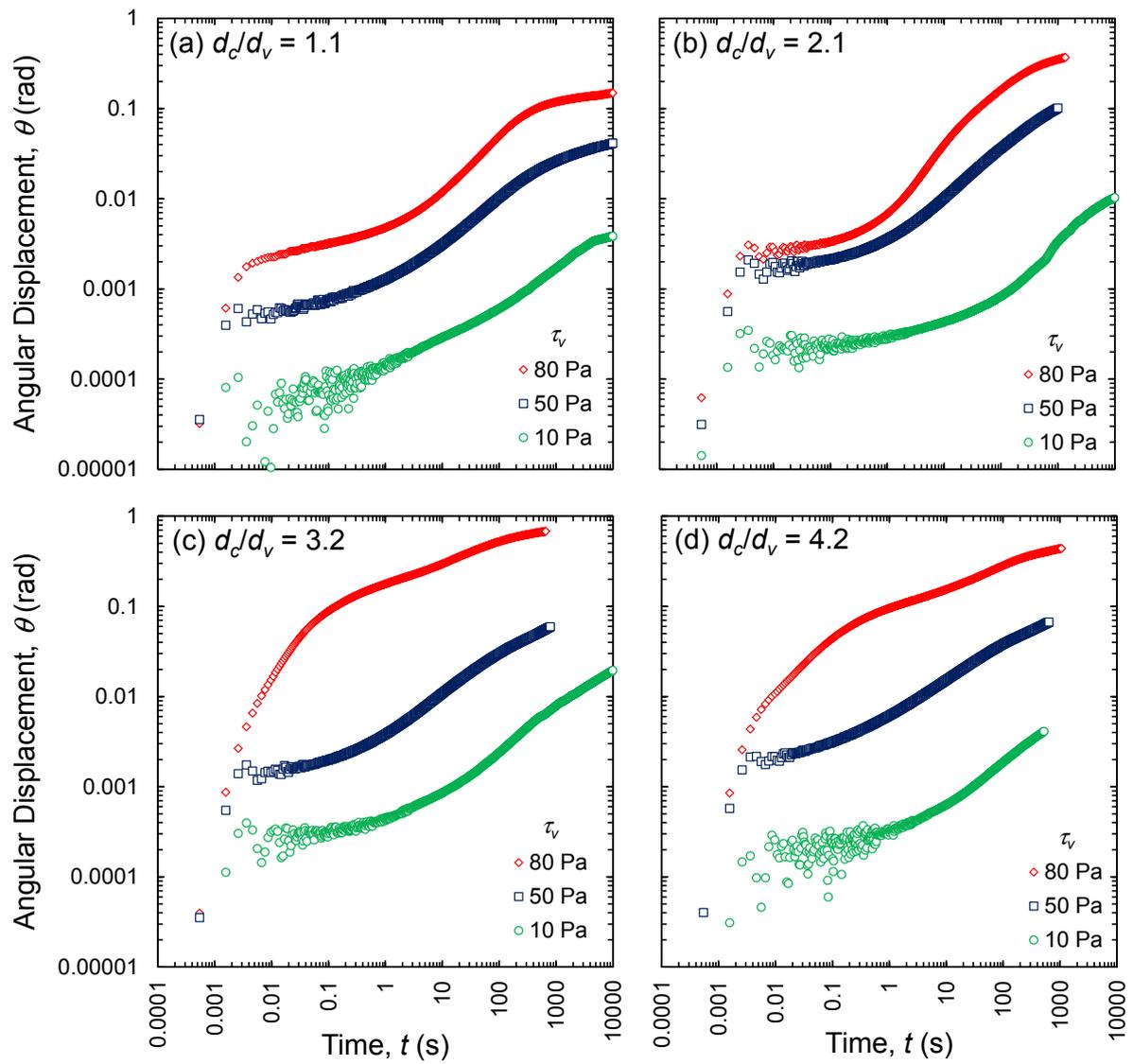

**Figure 14: Angular displacement as a function of time for creep testing of polymer flocculated alumina in four cup-to-vane diameter ratios**

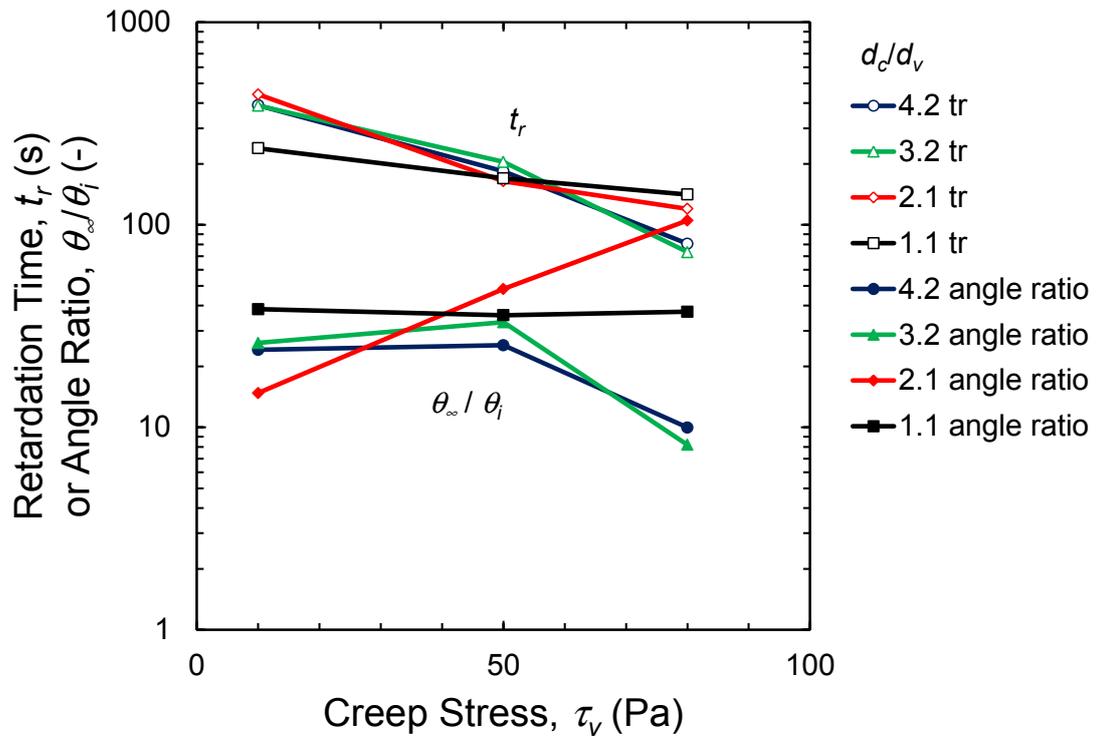

Figure 15: Retardation times and ratios of infinite to instantaneous creep angles for polymer-flocculated alumina in various cup-to-vane diameter ratios. $\theta_i = \theta(0.1\ s)$, $\theta_\infty = \theta(1000\ s)$, $\theta(t_{½}) = ½(\theta_i + \theta_\infty)$

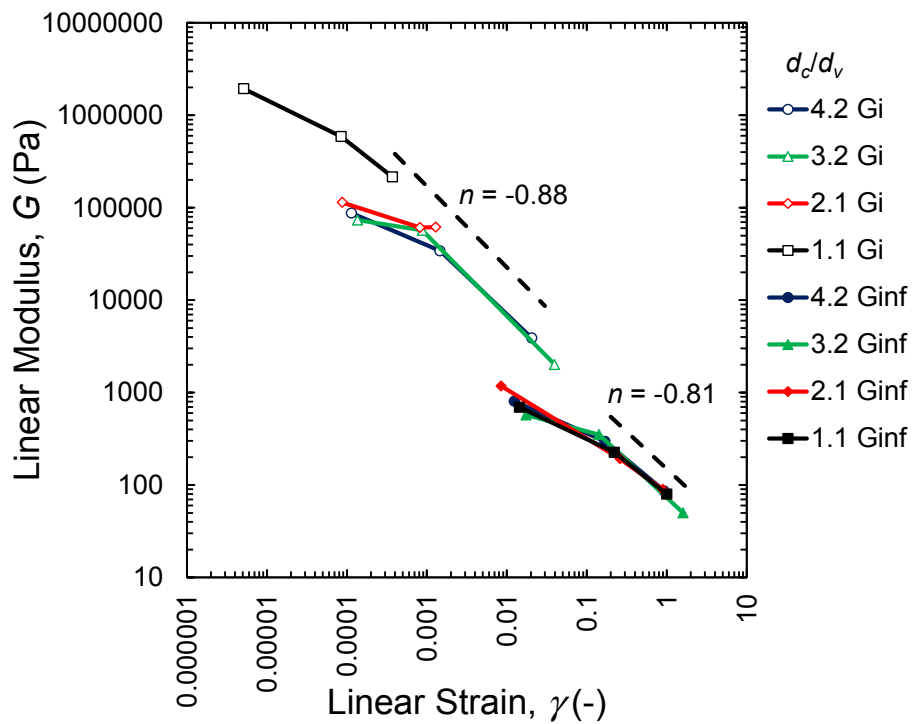

Figure 16: Instantaneous and infinite moduli and strain for polymer-flocculated alumina in various cup-to-vane diameter ratios. $G_i$ and $\gamma_i$ were calculated from $\theta(0.1\ s)$ while $G_\infty$ and $\gamma_\infty$ were calculated from $\theta(1000\ s)$, both using a linear approximation

## 5 Conclusions

Understanding the shear rheology of strongly-flocculated suspensions and improving current viscoplastic constitutive descriptions are important steps for improved prediction of suspension processing. In order to provide an enriched constitutive description, it is vital to have experimental data that are free from artefacts caused by the measurement geometry. The vane-in-infinite-cup geometry minimises wall effects by design. The aim of this work was to investigate the ratio of cup to vane diameter at which wall effects were minimised.

The results in this exploratory work showed consistent behaviour in larger diameter ratios across all three model systems, with linear viscoelasticity at low stresses, increased instantaneous compliance or non-linear strain softening at stresses closer to yielding, time-dependent yield over a range of stresses and finally instantaneous flow.

In contrast to the large diameter ratio results, the small diameter ratios showed a range of behaviours attributed to anomalous suspension behaviour at or near the cup wall. The coagulated titania and alumina at a diameter ratio of 1.1x showed yielding and subsequent slip at the wall, which indicated that the adhesive or wall yield stress was less than the cohesive or bulk yield stress. The apparent adhesive to cohesive yield stress ratios of 0.73 and 0.76 for the two coagulated suspensions here were used to estimate critical cup-to-vane diameter ratios of 1.17 and 1.15 to avoid wall slip. It is very important to realise though that the adhesive to cohesive yield stress ratio is likely to vary from one suspension to another and to depend upon the nature of the wall. For example, plastic walls might give lower adhesive to cohesive yield stress ratios and higher critical diameter ratios for aqueous mineral systems compared to metal walls, perhaps.

In addition to slip, the cup wall was observed to affect the pre-yield viscoelastic behaviour at cup-to-vane diameter ratios of 2.0x and 2.1x for the coagulated and polymer-flocculated alumina. In general, the suspensions appeared stiffer at stresses closer to yielding in the smaller diameter ratios. Analysis of the instantaneous and steady-state moduli for the coagulated alumina system using a modified Cross equation showed that the softening exponents decreased whereas the ratio of creep angles increased. Use of the small diameter ratio results to extract viscoelastic parameters ($G_i$, $G_r$ and $t_r$ from the SLS model as a minimum) would be incorrect by over an order of magnitude and mislead efforts to understand suspension rheology.

The strain softening behaviour of suspensions would not be expected to show constant softening index behaviour across all strains due to complex interactions at multiple length scales. For example, non-linear stretching and compression of inter-particle bonds could lead to strain hardening between the linear and softening regions. At higher deformations, rate-dependent viscous forces must also come into play. However, for a simple model with only three parameters, the modified Cross equation (especially the exact solution for $n = -1$) provided a good method of analysing the non-linear behaviour of the instantaneous and steady-state data.

The conclusion from this work was that artefact-free experimental results for studying the yielding behaviour of strongly flocculated suspensions required a vane-in-cup geometry with a minimum cup-to-vane diameter ratio of 3, although this value is conservative and must be dependent on the material. Based on this, existing creep results in the literature using a range of configurations with smaller diameter ratios and smooth tools may require reinterpretation.

Although premature or adhesive wall yielding and slip can be eliminated by means of wide gaps, their use is far from ideal in other regards as data analysis or inversion is then problematic, given non-linear viscoelasticity. It would be preferable arguably to use roughened cylinders and narrow gaps, although the problem then is to determine how rough the outer surface needs be to in order to cope with any material of interest. There is a need for more systematic work clearly, which could for example involve the use of splined outer cylinders in combination with a range of inner geometries, including a splined cylinder, a vane and a series of smooth cylinders of differing materials of construction. From data obtained using such it should then be possible to reconstruct what would be expected with a vane in wide and narrow smooth cylinders. More generally, the aim would be to determine the adhesive to cohesive yield stress ratio for a range of suspensions and materials in order to see how much it varies typically.

**Notation**

*Latin alphabet*

| | |
|---|---|
| $d$ | Diameter, m |
| $h$ | Height, m |
| $G$ | Modulus, Pa |
| $r$ | Radial position, m |
| $t$ | Time, s |

*Greek alphabet*

| | |
|---|---|
| $\gamma$ | Strain, - |
| $\theta$ | Angular displacement, rad |
| $\tau$ | Shear stress, Pa |
| $\omega$ | Rotational rate, rad/s |

Subscripts

| | |
|---|---|
| $c$ | Cup |
| $i$ | Instantaneous |
| $r$ | Retarded |
| $v$ | Vane |
| $\infty$ | Infinity |

**Acknowledgements**

The authors thank Daniel Lester for insightful discussions. Ashish Kumar was supported in this work by an Australian Postgraduate Award through the Australian Research Council. Tiara Kusuma was funded by a Melbourne International Research Scholarship through The University of Melbourne. Infrastructure support at Melbourne was provided by the Particulate Fluids Processing Centre, a Special Research Centre of the Australian Research Council. Simon Biggs acknowledges the support of the Nexia Solutions University Research Alliance for Particle Science and Technology and the University of Leeds. Simon Biggs and Amy Tindley were supported in this work as part of the TSEC programme KNOO and are grateful to the EPSRC for funding under grant EP/C549465/1.